\documentclass[11pt,twoside]{article}
\usepackage{macro-fsut-eng}
\usepackage{psfig}
\usepackage{graphicx}

\usepackage[T1]{fontenc} 

\usepackage{latexsym}
\usepackage{verbatim}

\begin{document}

\vskip 1.0cm
\markboth{M.S.~Nakwacki et al.}{Characterising the magnetic field in the intracluster medium}
\pagestyle{myheadings}

\vspace*{0.5cm}
\title{Characterizing the magnetic field in the intracluster medium}

\author{M. S. Nakwacki$^{1}$, E. M. de Gouveia Dal Pino$^1$, G. Kowal$^{1}$, R. Santos de Lima$^1$}
\affil{$^1$Instituto de Astronom\'ia, Geof\'isica e Ci\^encias Atmosf\'ericas, USP, 
Rua do Mat\~ao 1226, Cidade Universit\'aria, 05508-090 S\~ao Paulo, 
Brazil}

\begin{abstract}
During structure
formation, energetic events and random motions of the hot gas residing inside galaxy clusters (the intracluster medium, ICM) generate 
turbulent motions.  
Radio diffuse emission probes the presence of magnetic fields and
relativistic particles in the ICM, being a key ingredient for understanding the physical processes at work 
in clusters of galaxies. 
In this work, we present results from numerical simulations of magnetic field turbulence in 
magnetohydrodynamics (MHD) and kinetic MHD (KMHD) frameworks. We characterize 
magnetic field structures through the imprints left on polarization maps. 
\end{abstract}

\section{Results and conclusions}

We simulated MHD and KMHD turbulence in a 3D periodic box of $1$ Mpc$^3$ employing a modified Godunov-MHD code (Falceta-Gon\c alves et~al. 2008, Kowal et~al. 2011) with periodic boundaries, a resolution of $128^3$ grid points, ICM conditions (subsonic and superalfv\'enic), a 
gravity center mimicking the distribution of galaxies in a cluster, and a variable parallel to perpendicular pressure ratio  ($a_\parallel/ a_\perp$). We construct polarization maps ($RM$) using the simulated magnetic field and density. 
The KMHD model predicts a more granulated plasma density and magnetic field intensity, due to mirror
instability that accumulates energy in the smaller structures. As $a_\parallel/ a_\perp\rightarrow 1$ the gravity center leads to a more homogeneous density and magnetic field, probably due to an enhancement of particle mobility favoring their redistribution, the mixing  of magnetic field, and pressure isotropization. 
 The correlation between two different points can be measured by the structure function $S(r)$ (see e.g., Kowal et~al. 2011, for further details). 
From $S(r)$ we obtain a coherence length 
for the magnetic field $\sim 100$ kpc for KMHD (lower than the one obtained without a gravity center, e.g. $\sim 200$ kpc, Nakwacki et~al. 2012).

The gravity center decreases the magnetic field coherence length, indicating that its presence is needed to explain more accurately observational results.

\acknowledgments  We thank the Brazilian agencies FAPESP and CNPq.

\end{document}